\documentclass{ws-ijmpd}
\usepackage[super,compress]{cite}

\newcommand{\prd}{{\it Phys. Rev. D}}
\newcommand{\cqg}{{\it Class. Quantum Grav.}}

\begin{document}

\markboth{A. B. Balakin, A. E. Zayats}
{Nonminimal black holes with regular electric field}

\catchline{}{}{}{}{}
%
%%%%%%%%%%%%%%%%%%%%%%%%%%%%%%%%%%%%%%%%%%%%%%%%%%%%%%%%%%%%%%%%%%%%

\title{NONMINIMAL BLACK HOLES WITH REGULAR ELECTRIC FIELD}

\author{ALEXANDER B. BALAKIN}

\address{Department of
General Relativity and Gravitation, Institute of Physics, Kazan
Federal University, Kremlevskaya str. 18, Kazan 420008, Russia\\
Alexander.Balakin@kpfu.ru}

\author{ALEXEI E. ZAYATS}

\address{Department of
General Relativity and Gravitation, Institute of Physics, Kazan
Federal University, Kremlevskaya str. 18, Kazan 420008, Russia\\
Alexei.Zayats@kpfu.ru}

\maketitle

\begin{history}
\received{Day Month Year}
\revised{Day Month Year}
\end{history}

\begin{abstract}
We discuss the problem of identification of coupling constants, which describe interactions
between photons and space-time curvature, using exact regular solutions to the extended equations of the nonminimal
Einstein-Maxwell theory. We argue the idea that three nonminimal coupling constants in this theory can be reduced to
the single guiding parameter, which plays the role of nonminimal radius. We base our consideration on two examples of exact solutions obtained earlier in our works: the first of them
describes a nonminimal spherically symmetric object (star or black hole) with regular radial electric field; the second example represents a nonminimal Dirac-type object (monopole or black hole)
with regular metric. We demonstrate that one of the inflexion points of the regular metric function identifies a specific nonminimal radius, thus marking the
domain of dominance of nonminimal interactions.
\end{abstract}


\ccode{PACS numbers: 04.20.Jb, 04.40.Nr}

%\tableofcontents

\section{Introduction}

Nonminimal field theory, which takes into account specific interactions of scalar, pseudoscalar,
vector, electromagnetic and gauge fields with space-time curvature, extends significantly mathematical
and physical tools of theorists working in modern cosmology and astrophysics. For instance, the nonminimal coupling could explain
the Universe accelerated expansion, thus suggesting an alternative to introduction
of an exotic cosmic substrate, the dark energy (see, e.g., Ref.~\refcite{A1} and references therein). Nonminimal coupling of gauge, magnetic and electric fields to curvature
is shown to be able to support the stability of traversable wormholes, thus suggesting an alternative to exotic (phantom-type) sources (see, e.g., Refs.~\refcite{BSZ07} and \refcite{BLZ10}). One can say that
exact solutions to the nonminimally extended field equations form a very important supplement to the ``gold reserves'' of classical exact solutions obtained in the theory of gravity \cite{Exact}.
The main inconvenience of this Nonminimal field theory seems to be connected with a large number of phenomenologically introduced coupling constants, appeared in the Lagrangian of the theory. For instance,
the nonminimal Einstein-Maxwell theory contains three generic coupling parameters (see, e.g., Ref.~\refcite{BL05}); there are eight nonminimal coupling constants in the Einstein-Yang-Mills-Higgs theory (see, e.g., Ref.~\refcite{q2}); the nonminimal Einstein-Maxwell-axion theory includes nine coupling parameters (see, e.g., Ref.~\refcite{q3}).

Two important questions arise in this context. The first one is: how one can reduce the number of nonminimal coupling parameters to a minimum. The second question is the following: whether there exist a unique nonminimal coupling parameter with the dimensionality of radius, the {\it nonminimal radius}? We consider three methods to minimize the number of nonminimal coupling constants. First, one can use the requirements of regularity of the metric providing the following results: for spherically symmetric models we exclude distant singularities and eliminate the singularity in the center \cite{BZ07}; for the plane-wave symmetric models we require the metric determinant to be nonvanishing on infinite interval of retarded time \cite{q3}. The second method is to introduce geometric analogs for the tensor of nonminimal susceptibility. The third idea is to reduce new nonminimal parameters to the known coupling constants. These three methods are illustrated below in the framework of three-parameter nonminimal Einstein-Maxwell model (see Sec.~\ref{SecII}).

In this paper, we introduce the nonminimal radius of the spherically symmetric object, $r_{{\rm NM}}$,  by finding a boundary sphere, inside which the coupling of photons to curvature is the interaction of dominating type among ones forming gravity field. Mathematically, this radius can be found as a position of the corresponding inflexion point of the metric function $N(r)$, i.e., as the solution of the equation $N^{\prime \prime}(r_{{\rm NM}}){=}0$. In order to motivate such approach we consider, first, the zone of Coulombian attraction (repulsion) in the exact regular solution for the electric field obtained in Refs.~\refcite{BLZ10} and \refcite{BBL08} (see Sec.~\ref{SecIII}); in this illustrating example the inflexion point separates the Coulombian and nonminimal interaction zones.
Then we use the same idea searching for the inflexion points of the graphs for regular metric functions obtained in Ref.~\refcite{BBL08} for black holes with regular electric field (see Sec.~\ref{SecIII}), and in Refs.~\refcite{BZ07,BZ} and \refcite{BDZ1} for Dirac-type monopole (see Sec.~\ref{SecIV}). In these examples two inflexion points are visualized: the first one can be used for distinguishing the Newtonian-type attraction zone from the Reissner-Nordstr\"om-type zone of repulsion; the second inflexion point (closest to the center) indicates the domain of dominance of the nonminimal interactions.  Section~\ref{SecV} contains discussions and some prospects.

\section{Coupling constants in the nonminimal Einstein-Maxwell theory}\label{SecII}

\subsection{Reduction of the nonminimal susceptibility tensor}

The three-parameter nonminimal Einstein-Maxwell model can be
described in terms of the action functional
\begin{equation}\label{act}
S_{{\rm NMEM}} = \int d^4 x \sqrt{-g}\
\left[\frac{R}{8\pi G}+\frac{1}{2}F_{ik}
F^{ik}+\frac{1}{2} {\cal R}^{ikmn}F_{ik} F_{mn}
\right]\,,
\end{equation}
where, in addition to the standard elements: the Ricci scalar $R$ and Maxwell tensor $F_{ik}$, the interaction cross-term with the tensor of nonminimal
susceptibility ${\cal R}^{ikmn}$ is introduced. The tensor ${\cal R}^{ikmn}$ is given by the formula
\begin{align}
{\cal R}^{ikmn} &\equiv \frac{1}{2} q_1 R\,(g^{im}g^{kn} -
g^{in}g^{km}) \nonumber\\ {}&+ \frac{1}{2} q_2(R^{im}g^{kn} - R^{in}g^{km} +
R^{kn}g^{im} -R^{km}g^{in}) + q_3 R^{ikmn}\,, \label{sus}
\end{align}
where $R^{ik}$ and $R^{ikmn}$ are the Ricci and Riemann tensors,
respectively, and $q_1$, $q_2$, $q_3$ are the phenomenologically introduced coupling
parameters \cite{HO}. Generally, the parameters $q_1$, $q_2$, and $q_3$ are arbitrary
and have the dimensionality of length in square.
We think that the perfect nonminimal model has to contain only one fundamental parameter, say, $q$, and all three parameters $q_1$, $q_2$, $q_3$
should be expressed in terms of $q$ using some  linear relationships. As for the parameter $q$ itself, it can be either expressed in terms of well known fundamental constants,
or be nominated as new fundamental coupling constant. A few versions  are known, which realize this idea. The most known is the version of Drummond and Hathrell \cite{DH}, based on one-loop corrections to quantum electrodynamics in curved space-time, for which $q_1\equiv-5q$, $q_2=13q$, $q_3=-2q$, and the
positive parameter $q$ is constructed by using the fine structure constant $\alpha$, and the
Compton wavelength of the electron $\lambda_{\rm e}$, $q \equiv
\frac{\alpha\lambda^2_{\rm e}}{180\pi}$.

There are a few models, in which the linear relations between
$q_1$, $q_2$ and $q_3$ are proposed using some mathematical and/or physical motives, but the essential parameter $q$ is uncertain.
For instance, in Ref.~\refcite{MHZ} one can find the relations $q_1\equiv -q$, $q_2=2q$ and $q_3=-q$, which guarantee the equations for the
electric field potential $A_i$, and metric $g_{ik}$ to be of the second order in derivatives.
Keeping in mind that for this model (indicated as the Gauss-Bonnet model)  the nonminimal susceptibility tensor ${\cal R}^{ikmn}$ is
proportional to the double dual Riemann tensor $^{*}R^{*}_{ikmn}$, we suggested in Ref.~\refcite{BL05} to use geometric analogies to link coupling constants $q_1$, $q_2$, $q_3$
with $q$. For instance, one can use the condition  $^{*}{\cal R}^{ikmn} = {\cal R}^{*ikmn}$, which means that left-dual nonminimal susceptibility tensor coincides with the right-dual one. This condition provides only one relation $q_2+q_3=0$, and the second one is still necessary. As an alternative, we can consider the condition $g_{im}g_{kn}{\cal R}^{ikmn}=0$ (the trace of the nonminimal susceptibility tensor vanishes), which provides that $6q_1+3q_2+q_3=0$. Analogously, one can use the condition that ${\cal R}^{ikmn}$ is proportional to the difference between the Riemann and Weyl tensors, providing the relations: $q_1=-q$, $q_2=3q$ and $q_3=0$.

Another idea is to define $q_1$, $q_2$, $q_3$ so that the space-time metric, as the exact solution of the gravity field equations, is regular in the center.
For instance, the conditions $q_1=-q$, $q_2=4q$, $q_3=-6q$ guarantee (see Refs.~\refcite{BZ07,BZ} and \refcite{BDZ1}) that the nonminimal monopole of the Dirac type is regular everywhere.
With the conditions $q_1=-q$, $q_2=q$, $q_3=0$  (see Ref.~\refcite{BBL08}) and  $q_1=-q$, $q_2=3q$, $q_3=0$ (see Ref.~\refcite{BLZ10}) we obtained exact solutions characterized by the electric field
regular in the center. Thus, using physical and geometrical analogies, we can reduce the set of three nonminimal coupling parameters, $q_1$, $q_2$, $q_3$, to the single coupling constant $q$.

\subsection{Searching for the nonminimal radius}

Next problem with the identification of the nonminimal radius $r_{{\rm NM}}$ can be explained as follows. The simplest version is to consider $r_{{\rm NM}}=r_q$, where
the parameter $r_q$ with the dimensionality of length is defined using the nonminimal parameter $q$ only, $r_q=\sqrt{2|q|}$. In this sense
the Drummond-Hathrell radius $r_{\rm DH}$ is given by $r_{\rm DH} \equiv \lambda_{\rm e} \sqrt{\frac{\alpha}{90\pi}}$. But the definition $r_{{\rm NM}}=\sqrt{2|q|}$ is not unique.
In Ref.~\refcite{BZ07} we faced with the exact solution to the nonminimally extended Einstein-Maxwell equations, which gives the following metric function $N(r)=-1/g_{rr}$:
\begin{equation}\label{s1}
N(r)= 1 -\frac{2GM}{r} + \frac{GQ^2}{r^2} + \frac{2|q| GQ^2}{r^4} \,.
\end{equation}
This exact solution describes the nonminimal extension of the Reissner-Nordstr\"om solution, where $M$ is the asymptotic mass of the object, $Q$ is its total electric charge
(we use the system of units with $c=1$). In order to visualize three typical radii it seems to be reasonable to rewrite this metric function as
\begin{equation}\label{s2}
N(r)= 1 -\frac{r_{{\rm g}}}{r} + \frac{r^2_{{\rm Q}}}{r^2} + \frac{r^4_{{\rm NM}}}{r^4} \,,
\end{equation}
using the following definitions
\begin{equation}\label{s3}
r_{{\rm g}}\equiv 2GM \,, \quad r_{{\rm Q}}= \sqrt{G}|Q| \,, \quad r_{{\rm NM}} \equiv \left( 2|q| G Q^2\right)^{\frac14}\,.
\end{equation}
In this context the geometric mean $r_{{\rm NM}} {=} \sqrt{r_{q} r_{Q}}{=}\left( 2q G Q^2\right)^{\frac14}$ seems to be appropriate nonminimal radius.
Of course, a physically motivated choice of the nonminimal radius can differ from  $\sqrt{r_{q} r_{Q}}$ by some numerical coefficient.
In general case, using the model with arbitrary $q_1$, $q_2$, $q_3$ and the metric
\begin{equation}\label{s4}
ds^2 = \sigma^2(r)N(r) dt^2 - \frac{dr^2}{N(r)} - r^2\left(d\theta^2 + \sin^2{\theta} d\varphi^2 \right) \,,
\end{equation}
we obtained the following asymptotical decompositions for big values of the radial variable (see Ref.~\refcite{BLZ10}):
\begin{align}
\sigma(r)&=1+\frac{GQ^2}{2r^4}(10q_1+6q_2+3q_3)+\dots\,,\label{sel}\\
N(r)&=1-\frac{2GM}{r}+\frac{GQ^2}{r^2}-\frac{2GQ^2}{r^4}(4q_1+3q_2+2q_3)+\dots\,.\label{Nel}
\end{align}
These formulas demonstrate that the deviations of the nonminimal solutions from the minimal Reissner-Nordstr\"om metric functions start with the terms of the fourth order in $\frac{1}{r}$, and the  physically motivated value of radius $r_\text{NM}$ differs by the numerical coefficient $|\xi|$ only from the parameter $\sqrt[4]{2|q| GQ^2}$ introduced above for the particular model ($r_\text{NM} = |\xi| \sqrt[4]{2|q| GQ^2}$, where $4q_1+3q_2+2q_3 = \xi q$). The dimensionless parameter $\xi$ can be indicated as nonminimal fine structure parameter; below we discuss a few variants of identification of this quantity.

\section{Nonminimal model with regular electric field}\label{SecIII}

\subsection{Searching for a border surface distinguishing the Coulombian zone of attraction/repulsion}

Classical function describing the solution for the static electric field ${\cal E}(r)=\frac{Q}{r^2}$ (for sake of definiteness, below we consider $Q>0$) is infinite in the center ${\cal E}(0)=\infty$, has neither extrema, nor inflexion points, in which ${\cal E}^{\prime}(r)=0$ and ${\cal E}^{\prime \prime}(r)=0$, respectively. The situation changes principally, when we discuss exact solutions obtained in the framework of nonminimal Einstein-Maxwell theory.
In Ref.~\refcite{BBL08} for the case $q_1=-q$, $q_2=q$, $q_3=0$ (i.e., $|\xi|=1$), we obtain the following exact solution to the nonminimally extended Maxwell equations, describing static electric field of the spherically symmetric object
\begin{equation}\label{E1}
E(x) = \frac{Q}{2r^2_{Q}(1+x^2)} \left[1-x^2 + \sqrt{x^4+2x^2+5} \right] \,, \quad x = \frac{r}{r_{Q}} \,.
\end{equation}
This very illustrative exact solution is obtained for the specific model with $2q=r^2_{Q}$, i.e., when the parameter $r_q=\sqrt{2|q|}$ coincides with the electric radius $r_{Q}$. This solution has the standard Coulombian asymptotic behavior at $r \to \infty$:
\begin{equation}\label{E2}
E(x \to \infty) \sim  \frac{Q}{r^2_{Q}x^{2}} = \frac{Q}{r^2}  \,.
\end{equation}
In the center the electric field is regular and has a local maximum:
\begin{equation}\label{E3}
E(0) = \frac{Q}{r^2_{Q}} \cdot \frac{\left(1 + \sqrt{5} \right)}{2} \,, \quad E^{\prime}(0)=0 \,, \quad E^{\prime \prime}(0) = -\frac{Q}{r^2_{Q}} \cdot 2 \left(1+ \frac{2}{\sqrt5} \right) \,.
\end{equation}
The coefficient $\phi =\frac{\left(1 + \sqrt{5} \right)}{2}$ in the formula for $E(0)$ is equal to the golden section $\phi$, the limit to which the ratios of subsequent Fibonacci numbers tend.
Since the derivative
\begin{equation}\label{E4}
E^{\prime}(x) = - \frac{Q}{r^2_{Q}} \cdot \frac{2x}{(1+x^2)^2} \left[1+ \frac{2}{\sqrt{x^4+2x^2+5}} \right]
\end{equation}
has only one root at $x=0$ and is negative at $x>0$, the function $E(x)$ has only one extremum, the maximum in the center.
Since $E^{\prime}(0) = 0 = E^{\prime}(\infty)$, there is at least one inflexion point, $x=x_{*}$, in which $E^{\prime \prime}(x_{*})=0$. Numerical calculations show that this inflexion point is unique and is situated at $x_{*}=0.5528$. This means that we can indicate the sphere of the radius $r_{*}=r_{Q} x_{*}$ as the separatrix, which divides the 3-space into the zone of Coulombian-type interaction, defined as $r>r_{*}$, and the domain of nonminimal interaction, where $r<r_{*}$.

For arbitrary $q_1$, $q_2$, $q_2$ the corresponding solution is much more sophisticated, but for illustration of the idea we can use the following asymptotic decomposition of the electric field:
\begin{equation}
  E(r)=\frac{Q}{r^2}\cdot\left(1+\frac{4q_3GM}{r^3}-\frac{2GQ^2(q_2+3q_3)}{r^4}\right)+\dots\,.
  \label{E5}
\end{equation}
When $q_3=0$ and $q_2=q$, we face with the same nonminimal radius $r_{{\rm NM}} = \sqrt[4]{2|q| GQ^2}$. When $q_3 \neq 0$, formally speaking, the nonminimal radius can be defined alternatively as follows:
\begin{equation}
\tilde{r}_{{\rm NM}} = \sqrt[3]{|q_3|GM} = \sqrt[3]{\frac{|q_3|}{4|q|}} \cdot \sqrt[3]{ r^2_q r_{{\rm g}}} \,.
  \label{E6}
\end{equation}
Clearly, the radius $\tilde{r}_{{\rm NM}}$ can be indicated as {\it electric} nonminimal radius.

\subsection{Searching for the border surfaces distinguishing the Newtonian-type zone of attraction, the Reissner-Nordstr\"om-type zone of repulsion, and zone of nonminimal dominance}

Let us consider, first, the well-known minimal solution ($q=0$) for the massive electrically charged object with the metric of the Reissner-Nordsr\"om type:
\begin{equation}
\sigma(r) = 1 \,, \quad {\cal N}(r) =1 - \frac{2GM}{r} + \frac{GQ^2}{r^2} \,.
\label{RN1}
\end{equation}
The plot of this function has one extremum: the minimum at $r=r_{({\rm min})}= \frac{Q^2}{M}$, and one inflexion point at $r=r_{*}=\frac{3Q^2}{2M}$.
Thus, we can state that in this model the Newtonian-type attraction zone is situated at $r>r_{*}$. The violation of the Newtonian attraction law takes
place due to the electric contribution, which forms the zone of repulsion with infinite barrier near the center.

\begin{figure}[t]
\begin{center}
\begin{tabular}{c}
\includegraphics[height=6cm]{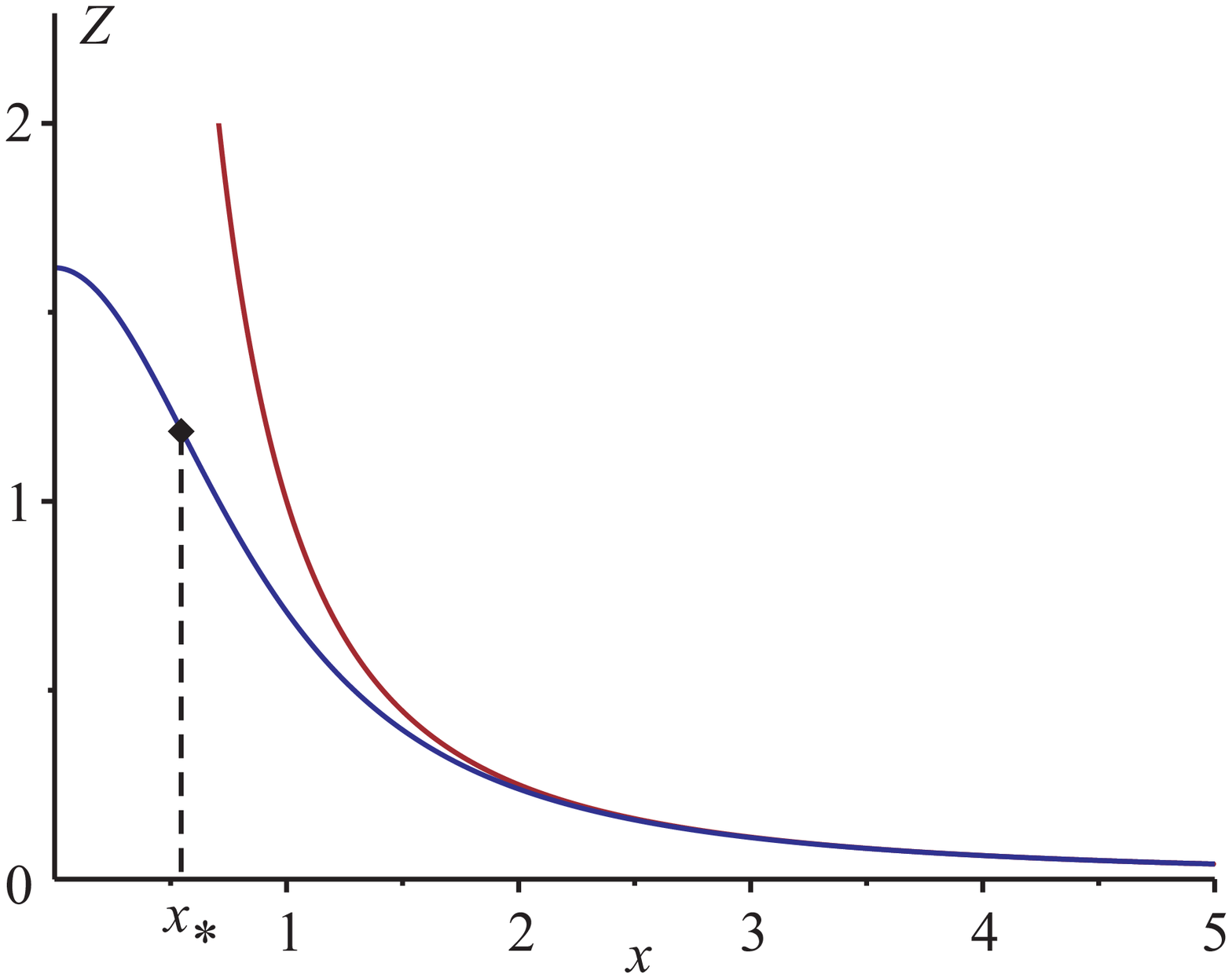} \\ (a) \\ \includegraphics[height=6cm]{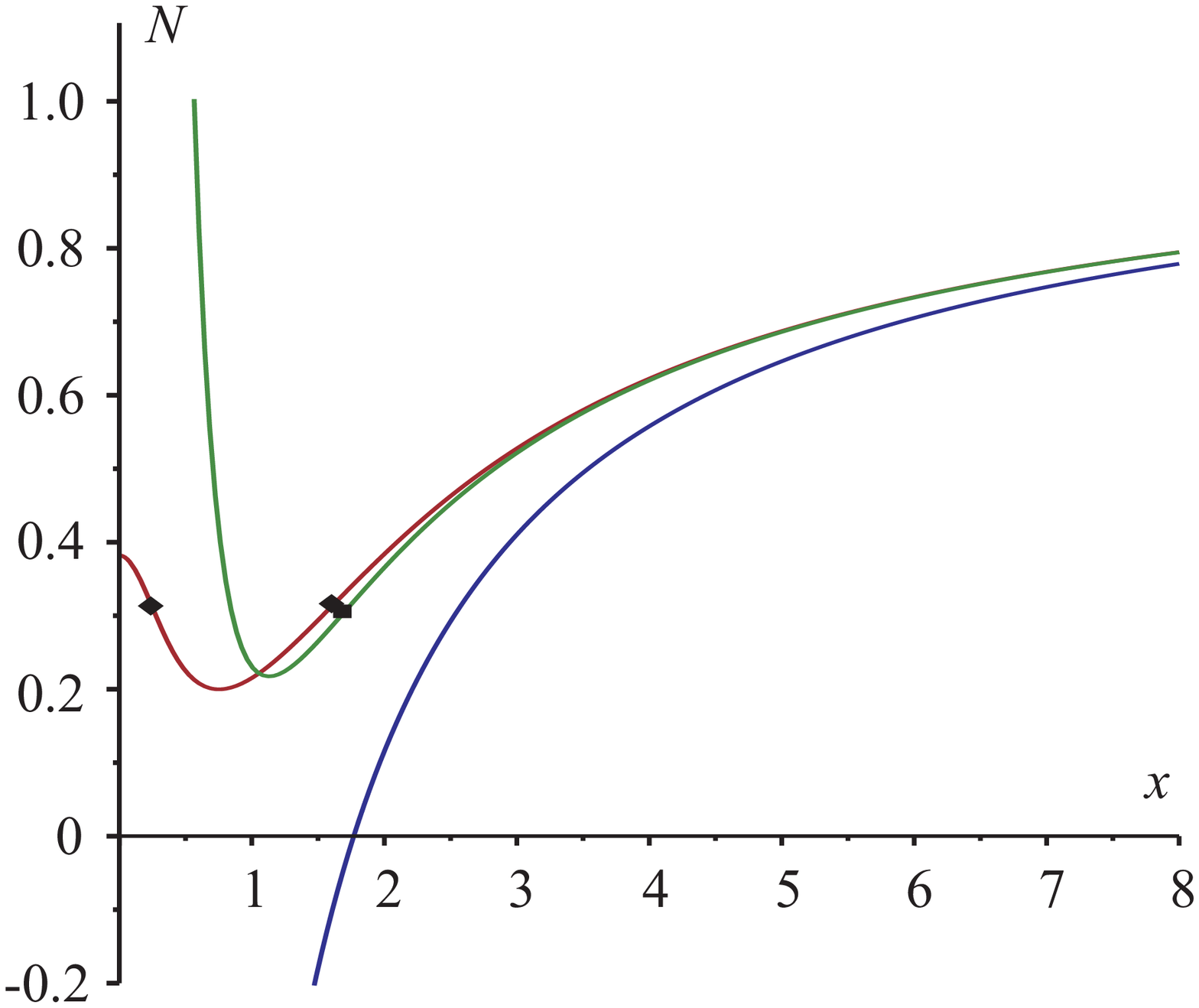}\\
(b)\\
\end{tabular}
\end{center}
\caption{ Panel (a) illustrates behavior of the function $Z(x)=E(r)r_Q/Q$, the dimensionless electric field, for two cases. The first curve relates to the Coulombian-type electric field infinite in the center; this function has no inflexion points. The second curve illustrates the electric field (\ref{E1}) regular in the center; it has the inflexion point $x_{*}=0.5528$, which marks the border surface distinguishing the zone of dominance of nonminimal interactions. Panel (b) contains three curves, which relate to the functions $N(x)$ for the Schwarzschild, Reissner-Nordstr\"om metrics, and for metric regular in the center, (\ref{E9}), respectively. The Schwarzschild function (in analogy to Coulombian-type solution) is infinite in the center and has neither extrema, nor inflexion points. The Reissner-Nordstr\"om curve is infinite in the center, but has one minimum and one inflexion point, which distinguishes the zone of Newtonian-type attraction. The third curve is regular in the center, has one minimum and two inflexion points; the inflexion point closest to the center marks the zone of dominance of nonminimal interactions.}\label{ElectN}
\end{figure}
When $q\neq 0$, the nonminimal coupling of photons to the space-time curvature provides the function $N(r)$ to be much more sophisticated. For the illustration, we consider the exact solution, which we presented in Ref.~\refcite{BBL08} for the case, when
$q_1=-q$, $q_2 = q$, $q_3=0$, $r_Q=r_\text{NM}$. The metric functions can be now written in the form
$$
\sigma(r)=\exp(-Z^2)\,,  \quad Z(x)=\frac{r^2_{Q}}{Q} \ E(x) \,, \quad  x=\frac{r}{r_Q} \,,
$$
\begin{equation}
N(r)=1+\frac{\exp(Z^2)}{x}\int\limits_0^x dx\,\exp(-Z^2)\left(2xZZ'(x)+x^2-Z^{-1}\right) \,.
\label{E9}
\end{equation}
In order to simplify the formulas, we used in (\ref{E9}) the term $E(x)$ from (\ref{E1}).
This exact solution relates to the critical value of the mass $M=M_\text{c}=0.8847 \frac{|Q|}{\sqrt{G}}$, and is characterized by the metric functions regular in the center:
\begin{equation}
N(0)=1-\frac{1}{Z(0)}=\frac{3-\sqrt5}{2}>0 \,.
\label{E10}
\end{equation}
Since $N(0)\neq 1$, the curvature invariants are infinite in the center, i.e., the solution has the so-called conical singularity in the center.
Numerical analysis shows that the function $\sigma(x)$ has only one inflexion point at $x^{(0)}_{*}=0.7773$.
The equation $N''(r_*)=0$ gives two inflexion points. The most distant point among them, $x^{(2)}_{*}=1.6137$, relates to the
border surface distinguishing the Newtonian-type zone of attraction. The first inflexion point, $x^{(1)}_{*}= 0.2489$, indicates the border surface
distinguishing the domain of dominance of nonminimal interactions. The zone between these two spheres can be indicated as the Reissner-Nordstr\"om-type zone (see panel (b) of Fig.~\ref{ElectN}).

\section{Regular nonminimal solution for the Dirac monopole of magnetic type}\label{SecIV}

Next example is the solution regular in the center, which we presented in Ref.~\refcite{BZ07}. This solution describes also the nonminimal Dirac monopole, when the mass $M$ is less than the critical one, $M_{{\rm c}}$, or the nonminimal black hole, when $M\geq M_{{\rm c}}$. The metric of such object can be represented as follows:
\begin{equation}
N(r) = 1+ \left(\frac{r_Q}{r_\text{NM}}\right)^2\cdot \frac{x^2(1-2mx)}{(1+x^4)} \,, \quad x = \frac{r}{r_{\text{NM}}} \,, \quad m = \frac{GMr_\text{NM}}{r^2_{Q}} \,, \quad \sigma(r)=1 \,.
\label{Reg1}
\end{equation}
Clearly, the function $N(r)$ depends on two effective guiding parameters: $m$ and $\frac{r_Q}{r_\text{NM}}$. In the center $N(0)=1$ for arbitrary pair of guiding parameters; thus, the metric is regular. Also, $N^{\prime}(0)=0$, i.e., there is a minimum in the center. The maximum exists for arbitrary $m$, but its height depends on $m$. Depending on $m$, a second minimum at $r=r_{{\rm min}}$ can appear; this minimum can relate, first, to the case $N(r_{{\rm min}})>0$ (regular nonminimal monopole); second, to the case $N(r_{{\rm min}})\leq0$ (regular nonminimal black hole); third, to the case $N(r_{{\rm min}})=0$ (regular nonminimal extremal black hole).

\begin{figure}[h]
%\begin{tabular}{cc}
\centerline{\includegraphics[height=6cm]{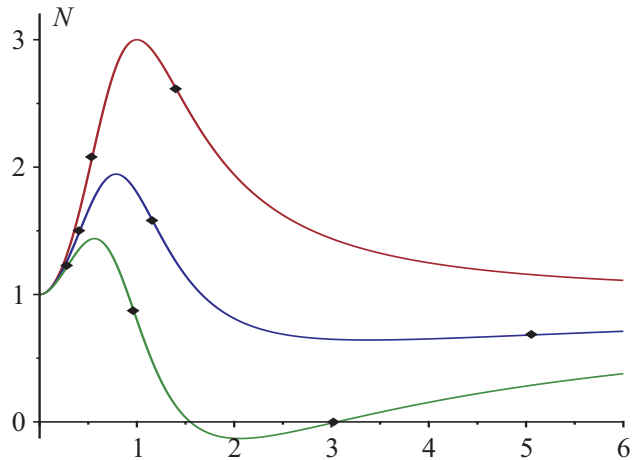}}%\qquad&\qquad\includegraphics[height=6cm]{NEE}\\
%a)&b)\\
%\end{tabular}
\caption{Plots of the metric function $N(x)$ given by (\ref{Reg1}) for $m=0$, $m=0.3$ and $m=0.55$, respectively. The upper curve, related to massless nonminimal Dirac monopole, has no Newtonian-type zone of attraction, and displays two inflexion points: the right one marks the zone of dominance of nonminimal interactions, and the left inflexion point distinguishes the trap near the regular center. The intermediate curve illustrates massive nonminimal Dirac monopole with three inflexion points. The bottom curve illustrates the solution for the nonminimal regular black hole; the curve has three inflexion points.}\label{MonN}
\end{figure}

We are interested in searching for the values of the radial variable $r=r_{*}$, for which $N^{\prime \prime}(r_{*})=0$, or in other words, the graph of the function $N(r)$ has inflexion points. To find these values  $r_{*}$, we have to solve the algebraic equation of the ninth order, which includes one guiding dimensionless parameter $m$. Instead of solutions $r_{*}(m)$ we prefer to analyze the reciprocal relation $m(r_{*})$, which is given by
\begin{equation}
m=\frac{3x_*^8-12x_*^4+1}{2x_*(x_*^8-12x_*^4+3)} \,.
\label{Reg2}
\end{equation}
Numerical analysis shows that in general there are three inflexion points in the graph of function $N(r)$.
The inflexion point $r^{(3)}_{*}$, the most distant from the center, lies in the interval $1.8512 r_\text{NM} < r^{(3)}_{*} < \frac{3Q^2}{2M}$; it indicates the
border surface distinguishing the Newtonian-type zone of attraction; clearly, at $r_\text{NM}=0$, when nonminimal coupling is switched off, the value of this parameter coincides with the value $\frac{3Q^2}{2M}$ obtained for the minimal Reissner-Nordstr\"om model. The inflexion point $r^{(1)}_{*}$, nearest to the center, belongs to the interval $0 < r^{(1)}_{*} < 0.5402\,r_\text{NM}$; it indicates the border surface distinguishing the nonminimal trap harboring the center. The intermediate inflexion point, $r^{(2)}_{*}$, relates to the interval $0.7109\,r_\text{NM} < r^{(2)}_{*} < 1.4066\,r_\text{NM}$; we consider namely this border surface as distinguishing the domain of dominance of nonminimal interactions. The appearance of three inflexion points are illustrated in Fig.~\ref{MonN}.

%\begin{figure}
%\includegraphics[height=6cm]{M}\caption{Plot of $x_*$ against $m=\frac{3x_*^8-12x_*^4+1}{2x_*(x_*^8-12x_*^4+3)}$}
%\end{figure}

\section{Conclusions}\label{SecV}

The nonminimal Einstein-Maxwell models considered in this paper illustrate explicitly the following ideas. The models involve three principal scales, which are described by
the Schwarzschild  radius $r_{{\rm g}}=2GM$, the Reissner-Nordstr\"om radius $r_Q=\sqrt{GQ^2}$ and the effective radius of nonminimal coupling, $r_{\text{NM}}$. The last parameter can be ambiguously determined, the variants being the following. The first variant relates to the simplest definition $\sqrt{2|q|}$, where $q$ the nonminimal parameter originated from the tensor of nonminimal susceptibility (\ref{sus}). The second version is connected with the value $\sqrt[4]{2 |q|G Q^2}$, which appeared naturally in the exact solutions obtained in Refs.~\refcite{BLZ10} and \refcite{BZ07}.
In addition, one can introduce the third variant, $\sqrt[3]{GM|q_3|}$, which relates to a new scale in the solution for electric field in the nonminimal model, which describes a deviation from the Coulombian law. When $q_3=0$, the deviation from the Coulombian law is described by the same scale parameter $r_{\text{NM}}=\sqrt[4]{2 |q|G Q^2}$.

For the further modeling it is important to rank three mentioned parameters, $r_{{\rm g}}$, $r_Q$ and $r_{\text{NM}}$. Our ansatz is that $r_Q < r_{\text{NM}}< r_{{\rm g}}$. This hypothesis is based on the estimations made with the Drummond-Hathrell model parameter $q= \frac{e^2\hbar}{180\pi m^2_e c^3}$, which guaranties that
$r_Q < r_\text{NM}=\sqrt[4]{2 |q|G Q^2}$, when $\frac{|Q|}{e}<10^{21}$.

Finally, we discussed a possibility to introduce a refined nonminimal parameter, say, $r_{{\rm R}}$, using the following finding. The curves, illustrating the behavior of regular electric field and of regular metric functions, display the presence of specific inflexion points, which distinguish the domain of dominance of nonminimal interactions. As it was shown above, the position of the corresponding border surfaces differ from $r_{\text{NM}}$ by some numerical coefficient of the order of one, e.g., 0.5402; 0.7109; 1.4066; 1.6137; etc. This finding can be interpreted as some supplementary motivation of the choice $r_{\text{NM}}= \sqrt[4]{2 |q|G Q^2}$ for the effective parameter of nonminimal coupling.

\section*{Acknowledgments}

Authors are grateful to Professor J.P.S. Lemos for fruitful discussions and advices. This work was supported by the Program of Competitive Growth
of KFU (Project 0615/06.15.02302.034), and by Russian Foundation for Basic Research (Grant RFBR N~14-02-00598).

\end{document}